\documentclass[preprint,showpacs,preprintnumbers,amsmath,amssymb,floatfix]{revtex4}

\usepackage{graphicx}
\usepackage{dcolumn}
\usepackage{bm}

\begin{document}

\preprint{APS/123-QED}

\title{Effect of the Casimir-Polder force on the collective 
oscillations of a trapped Bose-Einstein condensate}

\author{Mauro Antezza$^{a}$, Lev P. Pitaevskii$^{a,b}$ and Sandro
Stringari$^{a}$}
\email{antezza@science.unitn.it}
\affiliation{
$^{a}$Dipartimento di Fisica, Universit\`a di Trento
and Istituto Nazionale 
 per la Fisica della Materia INFM-BEC , I-38050 Povo, Trento, Italy \\
$^{b}$ Kapitza Institute for Physical Problems, ul. Kosygina 2, 117334 Moskow }

\date{\today}

\begin{abstract}
We calculate the  effect of the interaction between    an optically active material  and a Bose-Einstein condensate  on the collective oscillations of the condensate. We provide explicit expressions for the frequency shift of the center of mass oscillation  in terms of the potential generated by the substrate and of the density profile of the  gas. The form of the potential is discussed in details and various regimes (van der Waals-London, Casimir-Polder and  thermal regimes) are identified as a function of the distance of atoms from  the surface.  Numerical results for the frequency shifts are given for the case of a sapphire dielectric substrate interacting with a harmonically trapped condensate of $^{87}$Rb atoms. We find that at distances of $4-8 \mu m$, where thermal effects
become visible, the relative frequency shifts produced by the  substrate  are of the order $10^{-4}$ and  hence accessible experimentally. The effects of non linearities due to the finite amplitude of the oscillation are explicitly discussed. Predictions are also given for the radial breathing mode. 

\end{abstract}
\pacs{Valid PACS appear here}

\maketitle


\section{\label{sec:introduction}Introduction}

The study of the  force felt by an atom  near a surface has recently  become 
 a popular subject of research (see for example \cite{babb},\cite{milton} and references 
 therein) both in view of nanotechnological applications \cite{capasso} and  
 because of the possibility of investigating fundamental forces at the 
 submicron scale \cite{dimopulos}. These studies extend previous investigations 
 of the generalized van der Waals forces between macroscopic bodies \cite{24} to  more microscopic systems. Experiments with Bose-Einstein condensates near a surface  \cite{4,26,27} are very promising in this respect because of the high accuracy of the measurements achievable with ultracold gases.

Historically the  interatomic forces were first introduced in 1873 by van der Waals 
in order to explain the  deviations of the thermodynamic behaviour of real gases from  the ideal laws. Only in 1930 London \cite{london} provided a first  quantitative description of  the force. He applied second order perturbation theory to the quantum Hamiltonian to evaluate  the electrostatic interaction between two dipoles, induced by the fluctuation of the electromagnetic field.
He  succeeded in  deriving the most famous power law $1/d^6$, usually known as the van der Waals-London interaction, where $d$ is the interatomic distance.
Almost twenty years later, in 1948, Casimir and Polder \cite{CP},   taking into account relativistic retardation effects within fourth order perturbation theory, found that,  at  large distances, the potential decays like $1/d^7$ rather than like $1/d^6$.

The existence of these attractive forces between neutral atoms gives rise
 to analogous forces between a single atom and a dielectric or a metallic wall,
as well as to forces between macroscopic bodies separated by small distances
\cite{additivity}. Between the end of the fifties  and the beginning of 
sixties Lifshitz developed the general theory of the van der Waals forces
\cite{lifshitz,LP} for inhomogeneous media, taking into account  the quantum fluctuations of the electromagnetic field and their perturbation produced by the media. The theory describes the interaction between continuous media in terms of their dielectric functions and of the temperature. The van der Waals-London and Casimir-Polder forces are  recovered as limiting cases of the more general theory, which also accounts for the effects induced by the thermal fluctuations of the electromagnetic field. Thermal effects  provide the asymptotic behaviour of the interaction at distances much larger than the thermal wave length of photons.

The purpose of the present work is  to show that the study of the collective 
oscillations of a Bose-Einstein condensate of ultracold atoms can provide a 
useful  probe of such  forces.  Bose-Einstein condensates are
very dilute samples and consequently the force produced by the surface can be safely calculated starting
from the interaction felt by the individual atoms in the condensate \cite{additivity}. 
In this paper we will be 
mainly concerned with the center of mass oscillation of a condensate confined 
by a harmonic trap. The center of mass motion  
has the main advantage of being independent of  two-body  interactions and  
its frequency consequently provides an ideal probe of the additional  force 
generated by the wall. The effect of the force on the frequency of the center 
of mass oscillation depends on the distance of the condensate from the surface, 
on the temperature of the surface as well as on the optical properties of the 
surface and of the atoms in the condensate. 
The effect  also depends on the thickness of the condensate which in actual 
configurations can be of  the same order as the distance from the wall. 
In some limiting cases, corresponding to short and large distances, it is 
possible to obtain analytic or semi-analytic predictions that will be 
discussed systematically in our paper. They can serve as a natural guide 
for first quantitative estimates and as useful tests of more complete 
numerical approaches. 

The paper is organized as follows: 

In sect.2 we derive some general results for the shift of the  collective 
frequencies of a harmonically trapped Bose-Einstein condensate subject to 
an additional weak external force.

In sect.3 we  employ Lifchitz theory to calculate the force acting on 
a dilute atomic gas placed near a surface.

In sect. 4 we discuss  the transition from the  van der Waals-London potential, holding at very short distances from the surface and  decaying  like $1/d^3$, to the Casimir-Polder potential which accounts for relativistic retardation effects and holds at larger distances where it decays like $1/d^4$. 

In sect. 5 we discuss the transition from the $1/d^4$ law characterizing  the Casimir Polder potential  to the behaviour
of the potential at much larger distances  from the surface (larger than the thermal wavelength of photons) where it decays  like $1/d^3$ as a consequence of the thermal fluctuations of the electromagnetic field. 

In sect. 6 we discuss the role of the optical properties of the interacting media with  specific emphasis to the case of the dielectric sapphire ($Al_2O_3$) substrate and of the Rubidium atoms  which form the condensate. 

In sect. 7 we provide   results for the shifts of the collective frequencies as a function of the distance of the condensate from the surface, 
calculated at different temperatures.

Finally, in Appendix A1 we provide results for the frequency shift of the radial compression mode in very elongated harmonic traps.

\section{Center of mass oscillation of a trapped gas in the presence of a weak  perturbation.} 

After the first experimental realization of Bose-Einstein condensation in dilute vapours of alkali
atoms the experimental and theoretical research in the field of ultacold quantum gases has grown in an impressive way (for general reviews see, for example, \cite{rmp,leggett,CPHS,LPSS}). Bose-Einstein condensed gases are dilute, ultracold samples characterized by unique properties of coherence and superfluidity. They give rise, among others, to interference phenomena \cite{mitint,kasevich} as well as to a variety of collective oscillations \cite{SS,jila96,mit96}.  The relatively high density of Bose-Einstein condensed samples (compared to the one of non condensed trapped gases), as well as the possibility of imaging the atomic cloud after expansion, where the measured sizes become macroscopically large, permit to achieve very accurate determinations of the  frequencies of the collective oscillations, thereby providing interesting  opportunities for precision measurements. 

In this section we will calculate the effect of a weak  perturbation on the center of mass oscillation of a harmonically trapped gas.   The frequency of this oscillation, in the absence of the perturbation,  coincides with  the frequency of the  trap. This result, also known as Kohn theorem \cite{kohn}, is independent of the amplitude of the oscillation, interatomic forces, temperature as well as of the quantum  nature of the sample. Deviations of the measured frequency from the oscillator value can then be used as a useful probe of additional interactions, like the one generated by the surface. 
 In Fig.1 we show a schematic representation of the geometry considered in the present work. A dilute  gas trapped by the harmonic potential 
 \begin{equation}
V_{ho}({\bf r})={\frac{m}{2}}\omega_x^2x^2 + {\frac{m}{2}}\omega_y^2y^2 + {\frac{m}{2}}\omega_z^2z^2
\label{Vho}
\end{equation}
 is placed close to an ideal wall generating a potential $V_{surf}(z)$ so that the full potential felt by each atom is given by
 \begin{equation}
 V({\bf r}) = V_{ho}({\bf r}) + V_{surf}(z) \; .
 \label{V}
 \end{equation}
  In the following we will calculate the changes in the frequency of the center of mass oscillation along the $z$-th direction caused by the presence of the surface. In the appendix  we provide results also for the radial breathing mode
in the case of  very elongated axi-symmetric traps ($\omega_y \ll \omega_x=\omega_z$). 

The equations for the center of mass coordinate $Z_{CM} = (1/N)\langle \sum_{i=1}^Nz_i\rangle $ 
and the 
total momentum $P_z = \langle \sum_{i=1}^N p_{z_i}\rangle$ along the $z$-th direction ($N$ is the number of atoms of the trapped gas) can be 
written in the general form
\begin{equation} 
\frac{d Z_{CM}}{dt} = \frac{P_z}{Nm}
\label{Z}
\end{equation}
and 
\begin{equation} 
\frac{1}{N}\frac{d P_z}{dt}=  -{1\over N}\langle\sum_{i=1}^N \partial_{z_i} V({\bf r}_i)\rangle  = -m\omega^2_zZ_{CM} - \int d{\bf r} n({\bf r},t)\partial_zV_{surf}(z)
\label{P}
\end{equation}
where   we have used expression 
(\ref{Vho}) for $V_{ho}$ and we have expressed the average of the external force $\partial_z V_{surf}$ in terms of the density 
$n$  of the gas here assumed to be normalized to unity 
($ \int d{\bf r}n = 1$).
Notice that the equation for the total  momentum is not affected by  
the  two-body potential. 
In the absence of the perturbation ($V_{surf}=0$) one recovers  the equations of the harmonic oscillator and the 
center of mass oscillation corresponds to a rigid oscillation of the density. 
 The effect of the perturbation $V_{surf}$ can then be calculated, in  first order 
perturbation theory,  using the scaling ansatz  $n({\bf r},t) = n_0(x,y,z-Z_{CM}(t))$ in (\ref{P}),  where $n_0$ is the equilibrium density of the gas. The equation for the center of mass coordinate finally takes the simple form
\begin{equation}
{d^2 Z_{CM} \over dt^2} =  -\omega^2_zZ_{CM} - {1\over m}\int d{\bf r}n_0({\bf r}) \partial_z V_{surf}(z+Z_{CM})
\label{Psimple}
\end{equation}
where, in the integral, we have made the substitution $(z-Z_{CM}) \to z$. Equations (\ref{Z}) and 
(\ref{Psimple}) have now a closed form. In the limit of small oscillations we can expand the 
perturbation term up to terms linear in $Z_{CM}(t)$ and the equation for the center of mass takes again 
a harmonic  form  with the renormalized oscillator frequency 
\begin{equation}
\omega_{CM}^2 = \omega_z^2 +  \frac{1}{m}\;\int
 n_{0}({\bf r})\;\partial_z^2V_{surf}(z) \;d{\bf r}
\label{omegaD}
\end{equation}
It is worth stressing that result (\ref{omegaD}) is exact up to first order corrections in $V_{surf}$ and holds independently of the quantum or classical nature of the gas which instead determines the actual form of the density profile $n_0$.   Eq.(\ref{omegaD}) can be further simplified by introducing the so called 1D column density 
\begin{equation}
n^z_0(z) = \int dxdyn_0(x,y,z)
\label{n1}
\end{equation}
and using the fact that $V_{surf}$ depends only on the $z$-th coordinate. One then finds the result
\begin{equation}
\omega_{CM}^2 = \omega_z^2 + \frac{1}{m}\;\int_{-R_z}^{+R_z}
 n_{0}^{z}(z) \;\frac{d^2V_{surf}(z)}{dz^2}\;dz  \; .
\label{omegaD0}
\end{equation}

Equation (\ref{omegaD0}) shows that the ingredients needed to calculate the 
 effect of the perturbation on the shift of the center of mass frequency 
 are the explicit form of the  potential $V_{surf}$, which will be discussed 
 in the following sections, and the 1D column density (\ref{n1}).   
 For a Bose-Eintsein condensate in its ground state the density profile 
 is easily evaluated in the so called Thomas-Fermi approximation. This 
 approximation  holds for large condensates where, for gases interacting 
 with positive scattering length, the kinetic energy of the trapped condensate 
 can be neglected and one finds the analytic result \cite{rmp} 
 $n_0(x,y,z) =g^{-1}\;\left(\mu-V_{ho}({\bf r})\right)$ when $n_0\ge 0$ and $0$ elsewhere. Here 
 $\mu=(\hbar \omega_{ho}/2)(15Na/a_{ho})^{2/5}$ is the chemical potential,
  $N$ is the number of atoms, $\omega_{ho}=(\omega_x\omega_y\omega_z)^{1/3}$  is
  the geometrical average of the three trap frequences, $a_{ho} =
  \sqrt{\hbar/(m\omega_{ho})}$ is the oscillator length 
  and $g=4\pi \hbar^2 a/m$ is the interaction coupling constant fixed by 
 the s-wave 
 scattering length $a$. Integration of $n$ with respect to $x$ and $y$ yields
\begin{equation}
n_0^z(z) = {15 \over 16} {1\over R_z}\left(1-{z^2\over R^2_z}\right)^2
\label{n1BEC}
\end{equation}
where $R_z$ is the  Thomas-Fermi radius in the $z$-th direction, fixed by the relation
 $\mu = m \omega_z^2 R_z^2 / 2$. 
Typical values for $R_z$ in standard experimental conditions can be of the order of a few microns \cite{fermi}. 

Equations  (\ref{omegaD0}) and (\ref{n1BEC}) are the key result of this section and will be used in the second  part of the paper to calculate the effect of the  force produced by the surface on the frequency shift of the center of mass oscillation. 
The integral (\ref{omegaD0}) should be in general 
carried out numerically. 
There are, however,   important cases where the integration can be found   
analytically. This is the case, for example,  if one can approximate the 
potential with the algebraic decay $V^{(n)}(z)=-a_n/(d+z)^n$ where $d$ is the distance 
between the wall and the center of the harmonic trap \cite{sag}. In this case we find the result
 \begin{equation}\label{gamma4}
 \omega_{CM}^2 =\omega_z^2 -\frac{20\;a_4}{m\;d^6}\;
  \frac{1}{(1-\chi^2)^3}
 \end{equation}
  and
 \begin{equation}\label{gamma3}
 \omega_{CM}^2 =\omega_z^2 -\frac{15\;a_3}{4\;m\;d^5}
 \frac{\left[10\;\chi^3-6\;\chi-3(1-\chi^2)^2
 \;\ln\left((1-\chi)/(1+\chi)\right)\right]}{(1-\chi^2)^2\;\chi^5}.
 \end{equation}
for the most relevant $n=4$ and $n=3$ cases, respectively, where $\chi\equiv R_z/d$. 
Results (\ref{gamma4}) and (\ref{gamma3}) explicitly point out the role played by the finite size of the condensate.
In particular, only in the limit $R_z\ll d$ ($\chi \ll 1$) will the above equations approach the power law behaviours $1/d^6$ and $1/d^5$ for the frequency shift expected for a point-like condensate.

The above results  have been derived in the linear regime. If the amplitude of the oscillation
 is  comparable to the distance from the surface  non linear effects become important
  and modify  the value of the frequency shift.  
    Non linear effects result in a frequency shift as well
as in the occurrence of higher harmonics. They can be calculated starting from the general equations (\ref{Z}) and (\ref{Psimple}) for the center
of mass coordinate. Notice that, due to the harmonic nature of $V_{ho}$,  both effects
are absent 
if $V_{surf}=0$.  Due to the perturbative nature of the problem we can use the periodic law 
 $Z_{CM}= a\cos(\omega_{CM}t)$ to
evaluate  the integral of (\ref{Psimple}), where $a$ is the amplitude of the oscillation. One can then 
introduce the   periodic function  
\begin{equation}\label{3Lev}
Q(t)\equiv\frac{1}{m}\int d{\bf r} n_0({\bf r})\partial_zV_{surf}(z+a \cos(\omega_{CM}t))dz
\end{equation}
which can be conveniently expanded  in a Fourier series.
For the calculation of the frequency shift only the term proportional to
$\cos(\omega_{CM}t)$ is important and one can  write
\begin{equation}\label{5Lev}
\omega^2_{CM}-\omega_z^2 = \frac{\omega_{CM}}{\pi a}\int_0^{2\pi/\omega_{CM}}Q(t)\cos(\omega_{CM} t) dt.
\end{equation}
Expansion of (\ref{5Lev}) up to terms quadratic in $a$ and use of the 1D column density  finally yield the result
\begin{equation}\label{9Lev}
\omega^2_{CM}-\omega_z^2 =
\frac{1}{m}\int_{-R_z}^{R_z}n_0^z(z)\frac{d^2V_{surf}(z)}{dz^2}dz+\frac{a^2}{8\;m}\int_{-R_z}^{R_z}n_0^z(z)\frac{d^4V_{surf}(z)}{dz^4}dz
\end{equation}
for the frequency shift 
which generalizes eq.(\ref{omegaD0})  by including the non linear
correction in $a^2$.

\section{Force between the surface and a single atom.}

The interaction produced by a surface on a single atom includes, in its general form,  
non trivial relativistic effects as well as quantum and thermal fluctuations of the field.
  The force can be calculated starting from the general theory devoloped by
   Lifchitz 
  for the free energy associated with inhomogeneous media. For two plates separated by a distance $d$ and in thermal equilibrium with 
  the thermal radiation at temperature $T$, the force $F_{surf}$ per unit surface can be  written as \cite{LP}
\begin{gather} 
 F_{surf}(d)= \frac{k_B\;T}{16\;\pi\;d^3}\int_0^{\infty}x^2\bigg[
\frac{(\varepsilon_{10}+1)(\varepsilon_{20}+1)}{(\varepsilon_{10}-1)(\varepsilon_{20}-1)}e^x-1\bigg]^{-1}dx
+ \notag\\ 
+ \frac{k_B T}{\pi c^3}\sum_{n = 1}^{\infty} 
  \xi^{3}_{n}\int_{1}^{\infty} p^2 \bigg{\{}\bigg[\frac{(s_1 + p)(s_2 + p)}{(s_1 -
  p)(s_2 - p)}
\exp{\left(\frac{2\; p \;\xi_n \;d}{c}\right)}-1\bigg]^{-1}+ \label{surfacesurface}\\
+\bigg[\frac{(s_1 + p \;\varepsilon_1)(s_2 + p \;\varepsilon_2)}{(s_1 - p
\;\varepsilon_1)(s_2 - p \;\varepsilon_2)}\exp{\left(\frac{2\; p \;\xi_n\;
d}{c}\right)}-1\bigg]^{-1}\bigg{\}}\; dp\notag  
  \end{gather}
where, with respect to  the usual  presentation of the force, we have separated a first term behaving like $k_BT/d^3$ 
 from the other  terms with $n \ge 1$.
 In eq.(\ref{surfacesurface}) we have defined 
 \begin{equation} 
 s_1=\sqrt{\varepsilon_1-1+p^2} \;  , \;
s_2=\sqrt{\varepsilon_2-1+p^2}
\label{s1s2}
\end{equation} 
  where 
 $\varepsilon_1=\varepsilon_1(i\;\xi_n)$ 
and $\varepsilon_2=\varepsilon_2(i\;\xi_n)$ are the relevant dielectric functions of the two plates, 
evaluated for imaginary values $\omega=i\;\xi_n$ of the  frequency,  with 
 $\xi_n=2\pi k_B T n/\hbar$.
The function 
$\varepsilon(i\;\xi)$  is a monotonic and real function of $\xi$, obtained by analytic continuation 
of the non monotonic and complex  function $\varepsilon(\omega)$  evaluated for real values of the 
frequency $\omega$. Furthermore in eq.(\ref{surfacesurface}) we have introduced the static values
 of the dielectric functions $\varepsilon_{10}= \varepsilon_1(0)$ and 
 $\varepsilon_{20}=\varepsilon_2(0)$.
Some general properties of these  functions will be illustrated in sect VI. 

The force produced by a surface  on a single atom  can be easily derived by assuming that one of the two walls (wall 2) is made of a very dilute material (for example a gas) so that one can expand eq.(\ref{surfacesurface})
for small values of $\epsilon_2-1 =4\pi n\alpha$ where $n$ is the density of the gas and $\alpha$ is the corresponding  atomic polarizability. The potential $V_{surf}$ felt by a single atom is then simply given by $V_{surf}=-F_{surf}/n$ and can be written in the   form \cite{LP}:
\begin{gather} 
  V_{surf}(d)=-\frac{k_B\;T\;\alpha_{0}}
  {4\;d^3}\;\frac{(\varepsilon_{0}-1)}{(\varepsilon_{0}+1)}-\frac{K\;T}{ c^3}\sum_{n = 1}^{\infty}
   \alpha\;
  \xi^{3}_{n}\int_{1}^{\infty}  \exp{\left(-\frac{2\; p \;\xi_n \;d}{c}\right)}\;
   f(p)\; dp. \label{L}
  \end{gather}
where we have introduced the function $f(p)$  defined by 
\begin{equation} \label{fp}
  f(p)=\bigg[\frac{(s_{1} - p)}{(s_{1} + 
  p)} 
   +(1-2\;p^2)\frac{(s_{1} - p \;\varepsilon)}{(s_{1} +
   p\;\varepsilon)}\bigg]. 
\end{equation}
  In eqs.(\ref{L}) and (\ref{fp}) we have omitted the index $1$ in the dielectric function  characterizing  the optical properties of the surface while  $\varepsilon_{0}$ and $\alpha_0$ are the static values of the  dielectric
  function and of the atomic polarizability, respectively. 
Equation (\ref{L}) permits to describe both the interaction generated by 
a dielectric and by an ideal metal (in the latter case the function $\epsilon(\omega)$  behaves like $1/\omega$ at small $\omega$ and $(\epsilon_0-1)/(\epsilon_0+1)=1$.
 
Eqs. (\ref{surfacesurface}) and (\ref{L})  have been derived assuming that the whole system is in
thermodynamic equilibrium and in particular that  its components have
the same temperature $T$. In the experiments with Bose-Einstein
condensates the atomic gas is cooled down to extremely low
temperature, while the substrate remains at room temperature.  Result (\ref{L}) is
still expected to be valid in this case, $T$ being the temperature of the
substrate. In fact atoms, independently of being at zero or at room
temperature, do not contribute to the thermal radiation, their lowest
excitation energies being much higher than $k_BT$.

It is useful to identify the relevant length scales of the problem (see fig.2). A first important length is the 
thermal wavelength of the photon:
\begin{equation}
\lambda_T= {\hbar c \over k_BT}
\label{lambdaT}
\end{equation} 
where $k_B$ is the Boltzmann constant. At room temperature ($T=300 K$) one has $\lambda_T= 7.6 \mu m$. For distances $d$ larger than $\lambda_T$ the force is dominated by the thermal fluctuations of the electromagnetic field. In this limit the leading contribution  is given by the first term in eq.(\ref{L}), due to the occurrence of the exponential factor in the other terms,  and the potential takes the characteristic form 
\begin{equation} \label{highT}
V^T_{surf}(d)=-\frac{k_B\;T\;\alpha_{0}}
  {4\;d^3}\;\frac{(\varepsilon_{0}-1)}{(\varepsilon_{0}+1)}.
  \end{equation}
Notice that in this regime only the static value of the dielectric and polarizability functions contribute to the force. It is also worth noticing that in this limit
the  force is independent of the Planck constant as well as of the velocity of light. 

For distances $d$ smaller (or of the same order) than $\lambda_T$ the quantum fluctuations of the electromagnetic field become important and the terms of the sum (\ref{L}) should be taken into account.  If $d\ll \lambda_T$ and $T$ is sufficiently small, one can replace the sum with an integral (see sect.IV). In this regime one can introduce a second  length scale, hereafter called $\lambda_{opt}$.  This length  is related to  the relevant wavelengths characterizing the optical properties of the interacting media. One can identify two distinct regimes.   For distances $d\ll \lambda_{opt}$ the potential exhibits the familiar $1/d^3$  van der Waals-London dependence. In the interval 
$\lambda_{opt} \ll d \ll \lambda_T$ one instead enters the Casimir-Polder regime where the potential decays like 
$1/d^4$. The possibility of identifying the  Casimir-Polder  regime depends crucially on the value of the temperature. The temperature should be in fact sufficiently low in order to guarantee the condition
 $\lambda_T \gg \lambda_{opt}$ (see discussion at the end of sect. V). 
Differently from the thermal wawelength, given by the simple expression $\lambda_T$, the explicit evaluation of $\lambda_{opt}$ is not immediate because the optical properties of the media are not characterized by a single frequency (see sect.VI ). In the case of Rubidium atoms interacting with a dielectric sapphire substrate we find $\lambda_{opt} \sim 0.1 \mu m$. 

\section{The force at short distances:  from the van der Waals-London to the Casimir-Polder regime.}

As anticipated  in the preceding section at  distances smaller than the thermal wavelength
also the terms in the sum (\ref{L}) contribute to the force and, if $d\ll \lambda_T$ and the temperature is sufficiently low, the sum is conveniently 
 replaced by an integral: $\sum_n \to (\hbar/2\pi k_BT)\int d\xi$. 
 One then finds the useful expression 
\begin{equation}
V_{surf}^{SR}=-\frac{\hbar}{2\;\pi\;c^3}\int_0^{\infty}d\xi\int_1^{\infty}dp\;\;\xi^3\;\alpha\;
\exp{\Big(-\frac{2\;p\;d\;\xi}{c}\Big)}\;f(p),
\label{L0}
\end{equation}
for the potential, 
also called short range (SR) approximation. It is worth noticing that result (\ref{L0}) does not depend on
the value of $T$ \cite{Tneglected} and coincides with the zero temperature limit of the general form
(\ref{L}) of the potential.
Equation (\ref{L0}) admits two important limits. For large $d$ one finds the most famous
 expression
\begin{equation} \label{L0large}
V_{surf}^{CP}=-\frac{3\;\hbar\;c\;\alpha_{0}}{8\;\pi\;d^4}\;
\frac{(\varepsilon_{0}-1)}{(\varepsilon_{0}+1)}\phi(\varepsilon_{0}),
\end{equation}
also know as the Casimir-Polder law, where we have introduced  the function \cite{abrikosova}
  \begin{gather} 
    \phi(\epsilon) = \frac{(\varepsilon+1)}{(\varepsilon-1)}
    \bigg{\{}\frac{1}{3}+\varepsilon+\frac{4-(\varepsilon
    +1)\sqrt{\varepsilon}}{2\;(\varepsilon -1)} 
    -\frac{Arsh\sqrt{(\varepsilon - 1)}\;[1+\varepsilon+2\;\varepsilon
    (\varepsilon-1)^2]}{2\;(\varepsilon-1)^{3/2}}+ \notag\\
    +\frac{\varepsilon^2\;(
    Arsh\sqrt{\varepsilon}-Arsh\frac{1}{\sqrt{\varepsilon}})}
    {\sqrt{(\varepsilon+1)}}\bigg{\}}\label{phi}
    \end{gather}
which is equal to $1$  in the case of an ideal metal ($\varepsilon_0=\infty$). For sapphire ($\epsilon_0=9.4$) one finds
$\phi(\varepsilon_{0})=0.8$ 

In the opposit limit of small $d$ one instead finds 
the different law
  \begin{gather} \label{L0small}
 V_{surf}^{VL}=-\frac{\hbar}{4\;\pi\;d^3}\;\int_0^{\infty}\alpha(i\xi)\;\frac{\varepsilon(i\xi)-1}{\varepsilon(i\xi)+1}\;d\xi,
  \end{gather}
also know as  the van der Waals-London (VL) interaction. The coefficient characterizing this force is fixed by the optical properties
of the   media. The matching between the two laws (\ref{L0large}) and (\ref{L0small}) provides a useful estimate of the optical length
$\lambda_{opt}$ according to
\begin{equation}
\lambda_{opt}=\frac{3\;c\;\alpha_{0}}{2}\frac{(\varepsilon_{0}-1)}{\varepsilon_{0}+1}\phi(\varepsilon_{0})
\;\Big[\int_0^{\infty}\alpha(i\xi)\;\frac{\varepsilon(i\xi)-1}{\varepsilon(i\xi)+1}\;d\xi\Big]^{-1}.
\label{lambdaopt}
\end{equation} 
 In fig. \ref{fig:3} we show the  potential
(\ref{L0}) (dashed-dot line) calulated using the dielectric functions of Rubidium atom and of sapphire. The asymptotic laws $V^{VL}_{surf}=- 68.1/d^3 [nK (\mu m)^3]$ for the van der Waals   (see eq.(\ref{L0small})) and  $V_{surf}^{CP}= -8.34/d^4 [nK (\mu m)^4]$ for the Casimir-Polder  (see eq. (\ref{L0large})) potentials,   as well as the asymptotic thermal law $V_{surf}^T=- 2.86/d^3 [nK(\mu m)^3] $ evaluated at $T=300K$ (see eq.(\ref{highT})),  are also shown ($d$ is here expressed in microns). It is worth noticing that the potential (\ref{L0}) approaches the asymptotic Casimir-Polder force only at distances significantly larger than $\lambda_{opt}$.

\section{The force at large distances and the static approximation: quantum vs thermal effects.}

At  large distances from the wall the potential (\ref{L}) can be safely evaluated in the  so called static approximation (SA) which  consists of replacing  the dielectric functions $\varepsilon$ and $\alpha$ with their static values $\varepsilon_0$ and $\alpha_0$
 respectively. In the static approximation the  potential generated by the surface can be written in the 
useful form
 \begin{equation} \label{VG} 
  V_{surf}^{SA}(d)=-\frac{K\;T\;\alpha_{0}}{4\;d^3}\;
  \frac{(\varepsilon_{0}-1)}{(\varepsilon_{0}+1)}\;G(y)
  \end{equation}
where we have introduced the relevant scaling variable 
\begin{equation}
y=\frac{d}{\lambda_T}
\label{y}
\end{equation} 
and the  function 
\begin{equation}\label{G}
G(y)=\left[ 1+ 32\pi^3\;\frac{(\varepsilon_{0}+1)}{(\varepsilon_{0}-1)}
\int_{1}^{\infty}\;y^3\;g(4\;\pi\;p\;y)\;f_{0}(p) \; dp\right] \; .
\end{equation} 
In eq.(\ref{G}) $f_0(p)$ is the static limit of eq.(\ref{fp}), obtained by replacing $\varepsilon$ with
$\varepsilon_0$, while  the function $g$ is defined by 
\begin{equation}\label{17b}
g(a) \equiv \sum_{n=1}^{\infty}n^3\;e^{-a\;n}=\frac{e^{-a}\;\left[
1+4\;e^{-a}+e^{-2\;a}\right]}{(1-e^{-a})^4} \; .
\end{equation}
The asymptotic behaviours of the function $G(y)$ are well established: 
for $y\to 0$ one finds $G(y)\to
  3\phi(\varepsilon_{0})/(2\;\pi\;y)$.
Viceversa for $y\to \infty$ one finds $G(y) \to 1$. For a fixed value of $T$ the two 
limits apply, respectively,
to distances larger and smaller than the thermal wavelength (\ref{lambdaT}). It then
 follows that the static approximation (\ref{VG}) provides the transition from the 
 Casimir-Polder law (\ref{L0large}) holding for $\lambda_{opt}\ll d \ll \lambda_T$, to the thermal law 
 (\ref{highT}) holding for $d \gg \lambda_T$. In fig.4 we show the scaling function $G(y)$ together with the two 
 asymptotic limits  $y\to 0$ and $y\to \infty$  in the case of sapphire where we have 
 used the value  $\epsilon_0=9.4$. The function $G(y)$ depends rather smoothly on the actual value of the 
 static dielectric function, provided the  value $\varepsilon_{0}$ is significantly larger than $1$. 
 
 In fig. 3 we compare the prediction for the potential given by the
static approximation with the exact result evaluated starting from eq.(\ref{L}).
The comparison shows that for distances larger than $\sim 2 \mu m$ the
static approximation is rather accurate. This explains why the predictions
of this approximation for the frequency shifts are also accurate (see
figs \ref{fig:5} and \ref{fig:6}).

At shorter distances the static approximation fails and one should instead
use the short range description developed in sect.IV.  
If $\lambda_T \gg \lambda_{opt}$ the long range description based on the static approximation
 matches the  short range description of sect IV in the Casimir-Polder regime 
(\ref{L0large}). In this case the matching procedure completes the determination of the potential at
 all distances. If instead the condition $\lambda_T \gg \lambda_{opt}$
is not well satisfied 
 one should calculate explicitly the potential using the general 
expression (\ref{L}) and the potential will not exhibit the intermediate  Casimir-Polder 
behaviour (\ref{L0large}). For example in the case of  fig.3, where the potential is calculated at 
$T=300 K$, the matching between the two curves  is not very good and consequently the system never exhibits the Casimir-Polder behaviour \cite{opposite}. 
 
\section{Optical properties of the atom and of the substrate}

The functions $\epsilon(i\xi)$ and $\alpha(i\xi)$, which are crucial ingredients for the calculation of  the potential generated by the surface,
 obey the Kramers-Kronig relations
\begin{equation}\label{KK}
\varepsilon(i\;\xi_n)=
1+\frac{2}{\pi}\int_0^{\infty}\frac{\omega\;\varepsilon''(\omega)}{\omega^2+\xi_n^2}\;d\omega
\end{equation} 
where $\varepsilon^{\prime \prime}$ is the imaginary part of the dielectric function 
$\varepsilon
= \varepsilon^{\prime} + i \varepsilon^{\prime \prime}$ evaluated on the real axis. For 
the polarizability 
one has the analogous relationship
\begin{equation}\label{KK2}
\alpha(i\;\xi_n)=
\frac{2}{\pi}\int_0^{\infty}\frac{\omega\;\alpha''(\omega)}{\omega^2+\xi_n^2}\;d\omega
\end{equation} 
where  $\alpha^{\prime\prime}$ is the imaginary part of the atomic polarizability 
$\alpha = \alpha^{\prime} + i \alpha^{\prime \prime}$. 

The imaginary part of the dielectric function can be related to the measurable real an imaginary parts of the refraction index $n(\omega)= n^\prime(\omega) +
i n^{\prime \prime}(\omega)$
according the equation $\varepsilon''(\omega) = 2 n^\prime(\omega) n^{\prime \prime}(\omega)$. 
The values of $\epsilon^{\prime \prime}(\omega)$ (data taken from \cite{expsapphire}) and $\epsilon(i\xi)$ for sapphyire are reported on
 figs. \ref{fig:7} and \ref{fig:8}. From the optical point of view, crystalline sapphire is  an uniaxial crystal
 and one should distinguish between the   ordinary and extraordinary refraction indices.
 Since the optical properties for sapphire  are not completely known, in our calculations we have used only the data for ordinary waves. We expect that this approximation will not significantly affect the analysis in both crystalline and melted sapphires.
 In fig. \ref{fig:9} we  report the 
 corresponding values for   $\alpha(i\xi)$  for 
  $^{87}$Rb \cite{babbprivate}. The static value of the polarizability is $\alpha_0 = 47.3 \times 10^{-24} cm^{3}$.  Figure
\ref{fig:7} shows that the optical properties of sapphire are rather complex so that
 the actual value of $\lambda_{opt}$ governing the transition from the
  van der Waals-London to the Casimir-Polder regimes cannot be simply inferred
   from the form of $\epsilon^{\prime \prime}$, but requires the explicit
    calculation of the integral (\ref{lambdaopt}). The typical ``two plateau''
     behaviour exhibited by $\varepsilon(i\xi)$ in solid dielectrics is the consequence of the 
     concentration of the strength $\varepsilon^{\prime \prime}(\omega)$ in
      two distinct regions of frequencies (see figs \ref{fig:7} and \ref{fig:8}). 
      This behaviour is
       absent
in the atomic polarizability (see fig. \ref{fig:9}). Using eq. (\ref{lambdaopt}) it is
   possible to calculate the value of $\lambda_{opt}$. In the case of the 
 sapphire substrate interacting with Rubidium atoms we find 
 $\lambda_{opt} = 0.1 \mu m$. 

\section{Results for the shift of the center of mass oscillation}

We are now ready to calculate the shifts of the center of mass frequencies of the condensate caused by the  force generated by the surface. We will consider
a condensate of finite Thomas-Fermi radius $R_z$ located ad distance $d$ from the surface 
(see fig.1). We will evaluate
the fractional frequency shift 
\begin{equation}
\gamma  = {\omega_z-\omega_{CM} \over \omega_z} \; ,
\label{gamma}
\end{equation}
with $\omega_{CM}$ given by eq.(\ref{omegaD0}), as a function of the distance $d$  at different temperatures . As discussed in the preceding sections, the behaviour of the potential $V_{surf}$ at short distances is temperature independent and coincides with the zero temperature value. Conversely, at distances  of the order of the thermal wavelength, or larger, the potential exhibits an important temperature dependence. This behaviour is reflected in the shift of the center of mass frequency of the condensate for which we provide our predictions in fig. 10 at different temperatures. The curves of this figure have been calculating using the general espression (\ref{L}) for the potential. 
 In the calculation we have used the values $\omega /2\pi = 220 Hz$ for the harmonic trapping frequency along the $z$-th direction and $R_z= 2.5 \mu m$ for the $z$th radius of the condensate. These are typical values employed  in current experiments with Bose-Einstein condensates. We have used a sapphire substrate and the condensate is made of $^{87}$Rb atoms. Our predictions show that  
at distances $d \sim 4-8 \mu m$, where the thermal effects  start becoming 
important at room temperature, the frequency shifts are of the order  
of $10^{-4}$.  These results are promising in view of the possibility of systematic measurements of the interplay between quantum and thermal effects in the generalized van der Waals
forces.

It is also interesting to  compare  the predictions for the shifts of the 
collective frequencies with the 
results obtainined using the static approximation,  holding at large 
distances (see sect.V).  The comparison (see figs. \ref{fig:5} and \ref{fig:6}) shows that the static approximation  
(\ref{VG}) provides an  accurate decription of the shifts   for distances 
larger than 
$4\mu m$, even at zero temperature. At shorter distances the deviations become more important. One should actually
notice that at short distances the width 
of 
the condensate plays an important role amplifying the effects of the short range 
component of the potential.

Let us finally discuss the effects of non linearities on the frequency shift.
 If the amplitude of the oscillation is not small additional corrections due to 
 the external force should be  taken into account (see sec.II). The 
 corresponding corrections  can be estimated using result (\ref{9Lev}). 
 For example, by choosing $a= 0.5 \mu m$ and making the same choices for the other parameters ($R_z=2.5$ and $\omega_z/2\pi=220 Hz$), 
 we find that the prediction for the frequency shift (\ref{gamma})  is increased, in absolute value,   by $\sim 20 \%$ at $d=6 \mu m$. This result shows that for larger choices of the amplitude of the oscillation and for smaller values of $d$ the effects of non linearity can become very important. In general they should be calculated  starting directly from eq.(\ref{3Lev}) and (\ref{5Lev}).
 
 \section{Conclusions}
 
In this work we have applied Lifshitz theory of generalized van der Waals interactions to investigate the effect of a substrate on the collective oscillations of a trapped quantum gas.  We have first developed the general theory for  the frequency shifts of the center of mass oscillation caused by a perturbative force of general form. We have hence characterized the potential generated by the surface identifying various relevant regimes. A first approximation, holding at short distances $d$ from the surface, permits to describe the transition from the van der Waals-London $1/d^3$ law to the Casimir-Polder $1/d^4$ regime characterized by quantum and relativistic effects. A second approximation, holding at larger distances, permits to describe the transition from the Casimir-Polder law to the large distance $k_BT/d^3$ behaviour, dominated  by the thermal fluctuations of the electromagnetic field. We have also identified the relevant length scales of the problem and explored the matching conditions for the various approximations. Due to the finite size of the condensate the calculation of the frequency shifts requires a numerical integration, although analytical expressions can be obtained in some limiting cases. The  calculation  requires the explicit knowledge of the optical properties of the interacting media and in particular the
dielctric function  of the susbstrate and the atomic polarizability of the atoms in the condensate.  We have also exploited the effects of nonlinearity and shown that, due to the finite width of the condensate, they can provide large corrections to the frequency shifts and should be consequently   taken into account in the comparison with future experiments. We have finally  derived (see the Appendix) results  for the frequency shifts of the radial breathing mode. 

Our predictions are rather promising and suggest that Bose-Einstein condensates can become efficient sensors of very weak forces. In particular the oscillation of the condensate permit to explore the interplay between quantum and thermal fluctuations of the electromagnetic field in the relevant region of  $4-8 \mu m$. For such distances   the relative shifts are predicted to be of the order of $10^{-4}$ and are hence  measurable experimentally.

\section{Acknowledgements} 

This work was stimulated by insightful discussions with  Eric Cornell. It is a pleasure to thank him as well as 
John Obrecht, Jeffrey McGuirk and David Murray Harber
 for many  useful comments. 
We would also like to thank J.F. Babb for sending us the calculated values of $\alpha(i\xi)$ for  Rubidium. 

\section{Appendix: Frequency shift of the radial breathing mode.}

The radial breathing oscillation  in elongated traps  is a 
fundamental mode exhibited by  Bose-Einstein condensates. 
  For an interacting
Bose-Einstein condensate, harmonically trapped in two dimensions, this mode  exactly occurs  at twice the trapping radial frequency (we assume here symmetric trapping: $\omega_x=\omega_z \equiv \omega_{\perp}$),  
irrespective of the amplitude of the oscillation, number of atoms and  
value of the scattering length  \cite{lev}.  In a 3D configuration this mode can be realized 
in the limit of a very elongated axi-symmetric trap ($\omega_y \ll \omega_{\perp}$) corresponding, 
in the geometry of fig. 1,  to a condensate very elongated along 
 the   direction orthogonal to the figure. The radial breathing  oscillation 
has been measured with high accuracy in \cite{dalibard}. 

The radial breathing  mode  can be described by deriving the time-dependent equations for the average square radius
 $\langle x^2+z^2\rangle$  within 
  Gross-Pitaevskii theory \cite{LPSS} in 2D. After some straightforward algebra one finds the equation
 \begin{equation} 
{d^2 \over dt^2}\langle x^2+z^2\rangle = - {2\over m} \langle {\bf r}\cdot \partial_{\bf r} V\rangle +{2\over m^2}\langle p^2_x+p^2_z\rangle
+ {4\over m}\langle V_{2-body}\rangle  
\label{rr}
\end{equation}
where  $ V_{2-body}$ is the average values of the mean field interaction energy and $V$ is the external potential. By introducing the  energy per particle $E/N=\langle V\rangle + \langle p^2_x+p^2_z\rangle/2m+ \langle V_{2-body}\rangle$, which is conserved in time, and separating in $V$ the axi-symmetric harmonic potential $V_{ho}$   from the surface potential $V_{surf}$, eq.(\ref{rr}) can be rewritten as:
\begin{equation}
{d^2 \over dt^2}\langle x^2+z^2\rangle =  {4\over m}E -4\omega^2_{\perp}\langle x^2+z^2\rangle -{4\over m}\langle V_{surf}\rangle -  
{2\over m} \langle z \partial_z V_{surf}\rangle
\label{rr1}
\end{equation} 
In the absence of the surface perturbation ($V_{surf}=0$)  eq.(\ref{rr1}) gives oscillating solutions 
with $\omega=2\omega_{\perp}$. 

The effect of the perturbation is simply calculated noticing that the unperturbed 
solution corresponds  to a scaling transformation where the density varies as
$n(x,y,t)= \alpha^2n_0(x/\alpha,y/\alpha)$ and $n_0$ is the equilibrium density distribution. The time dependent scaling parameter $\alpha$ fixes the value of the square radius according to $\langle x^2+z^2\rangle = \alpha^2\langle x^2+z^2\rangle_0$. 
 Using the scaling transformation to evaluate the integrals $\langle V_{surf}\rangle = \int d{\bf r}n(x,y,t)V_{surf}(z)$
 and $\langle z\partial_zV_{surf}\rangle = \int d{\bf r}n(x,y,t)z\partial_zV_{surf}(z)$ one can rewrite eq.(\ref{rr1}) as  a simple equation for the parameter $\alpha$.  In the following we will be interested in the limit of small amplitude oscillations. By expanding the integrals up to 
first order  in  $(\alpha-1)$, one finally obtains the result 
\begin{equation}
\omega_{B}^2 = 4\omega_{\perp}^2  +{1\over m \langle x^2+z^2\rangle}\left[\int_{-R_z}^{R_z}dz n_0^z z\partial_z z \partial_z V_{surf} +2\int_{-R_z}^{R_z}dz n_0^z  z \partial_z V_{surf}\right]
\label{B}
\end{equation}
for 
the frequency shift of the radial breathing (B) mode, where $n_0^z$ is the 1D column density (\ref{n1}).
Similarly to the case of the shift of the center of mass frequency (see eq.(\ref{gamma})), 
also result (\ref{B}) is exact up to first order corrections in the perturbation. In fact  the radial scaling ansatz is 
an exact solution of the Gross-Pitaevskii equation in 2D. It is also a good approximation in 3D if the harmonic trap is sufficiently elongated.
Result (\ref{B}) for the radial breathing mode  may provide a further tool to investigate the effects of the generalized van der Waals force on a Bose-Einstein condensate.

\clearpage
\begin{figure}[ptb]
\begin{center}
\includegraphics[width=1.0\textwidth]{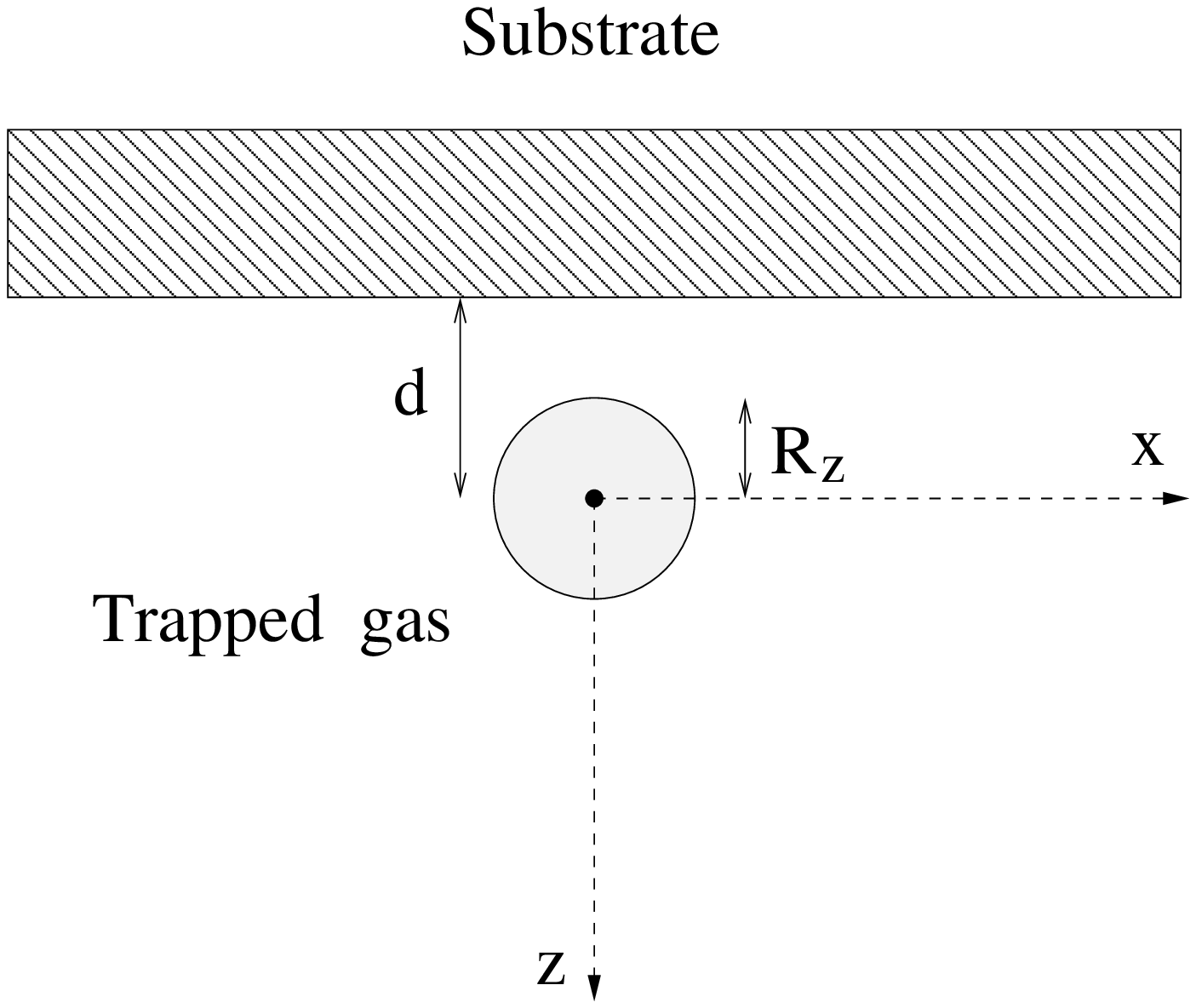}
\caption{\footnotesize Schematic figure of the substrate-trapped gas system. Gravity is oriented along $z$.}\label{fig:1}
\end{center}
\end{figure}
\clearpage
\begin{figure}[ptb]
\begin{center}
\includegraphics[width=1.0\textwidth]{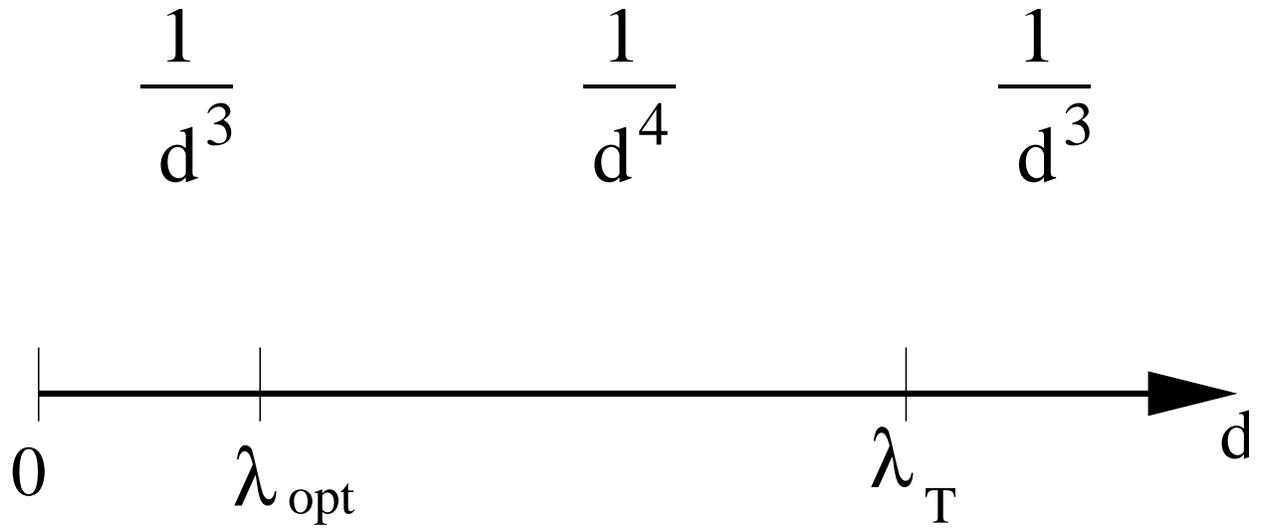}
\caption{\footnotesize Relevant length scales $\lambda_{opt}$ and $\lambda_T$ 
caracterizing the generalized van der Waals 
potential $V_{surf}$. The power law behaviour of $V_{surf}(d)$ is shown in the appropriate
ranges of validity as a function of the distance from the surface.}\label{fig:2}
\end{center}
\end{figure}
\begin{figure}[ptb]
\begin{center}
\includegraphics[width=1.0\textwidth]{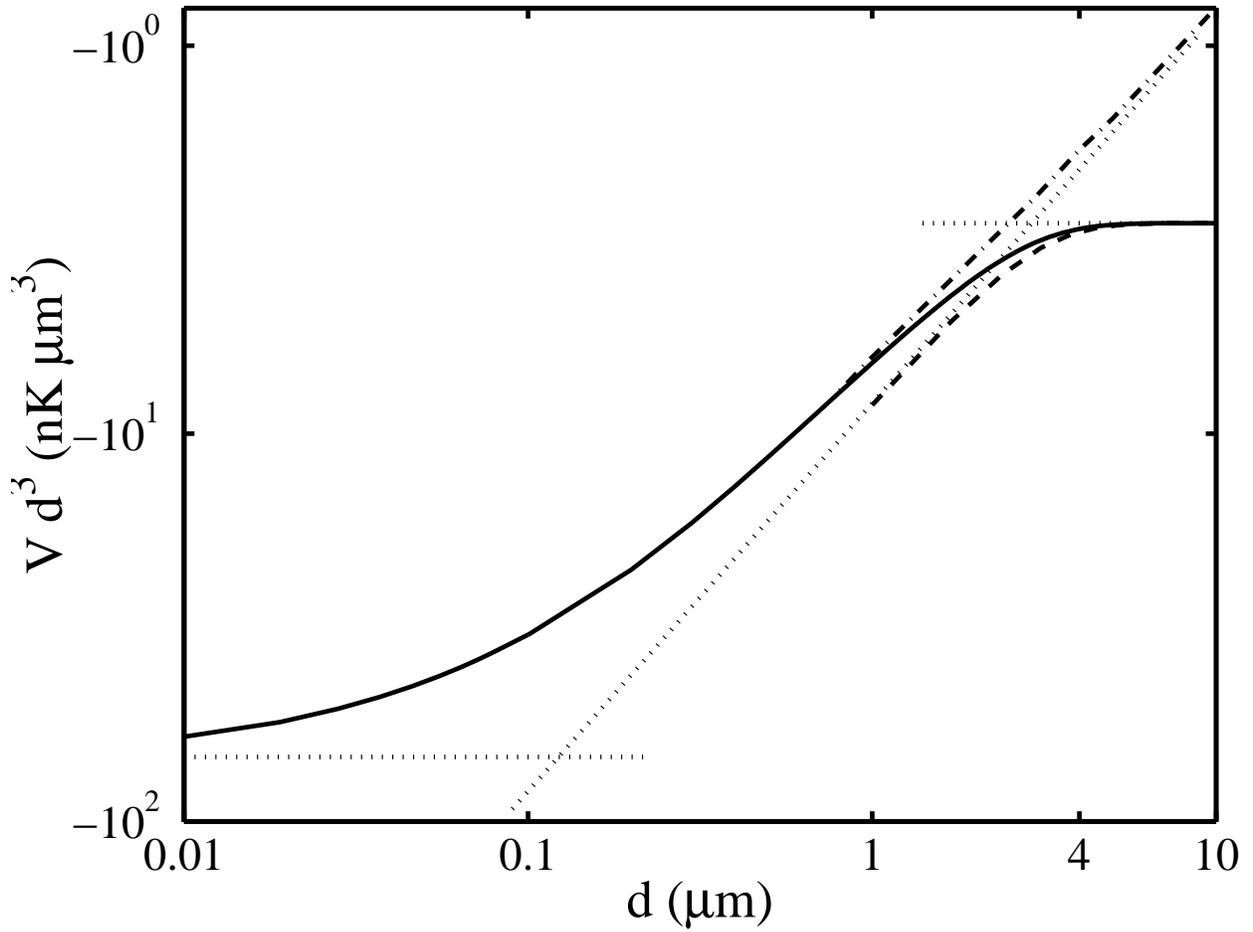}
\caption{\footnotesize The  atom-surface potential  is shown using the exact formula 
of eq.(\ref{L}) (solid line), the short range approximation (\ref{L0}) (dash-dotted line) and the
static approximation (\ref{VG}) (dashed line). The asymptotic van der Waals-London ($\sim 1/d^3$), 
Casimir-Polder ($\sim 1/d^4$)  and  high $T$ ($\sim 1/d^4$) potentials are also shown
(dotted lines). The curves were obtained for a sapphire substrate at $300 K$ and for  $^{87}$Rb atoms in the condensate.}\label{fig:3}
\end{center}
\end{figure}
\begin{figure}[ptb]
\begin{center}
\includegraphics[width=1.0\textwidth]{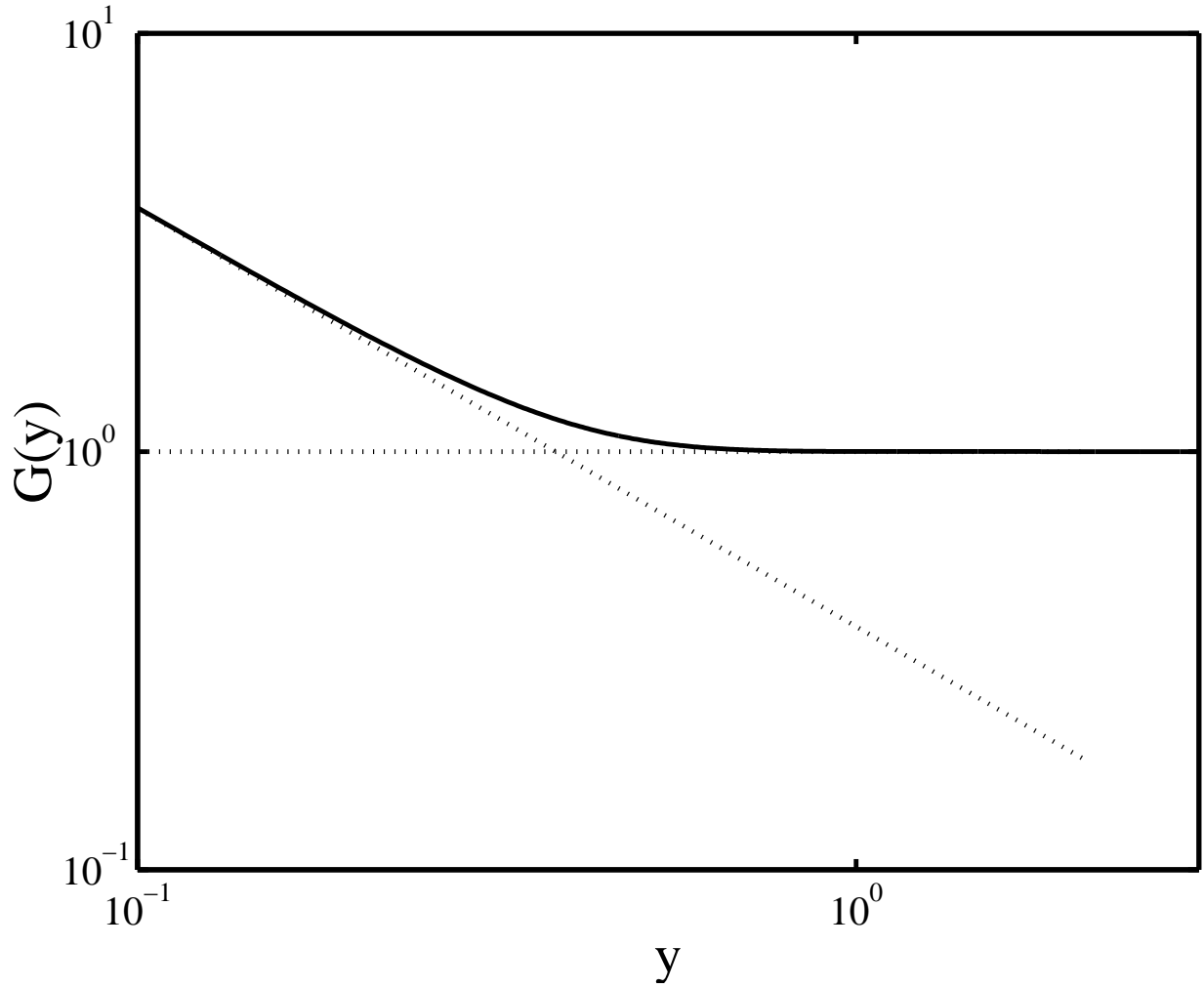}
\caption{\footnotesize The function $G(y)$  (solid line) is shown   as a function of the scaling variable
 $y=d/\lambda_T$ together with  its  asymptotic behaviours (dotted lines). The function was calculated for the sapphire substrate (see text).}\label{fig:4}
\end{center}
\end{figure}
\begin{figure}[ptb]
\begin{center}
\includegraphics[width=1.0\textwidth]{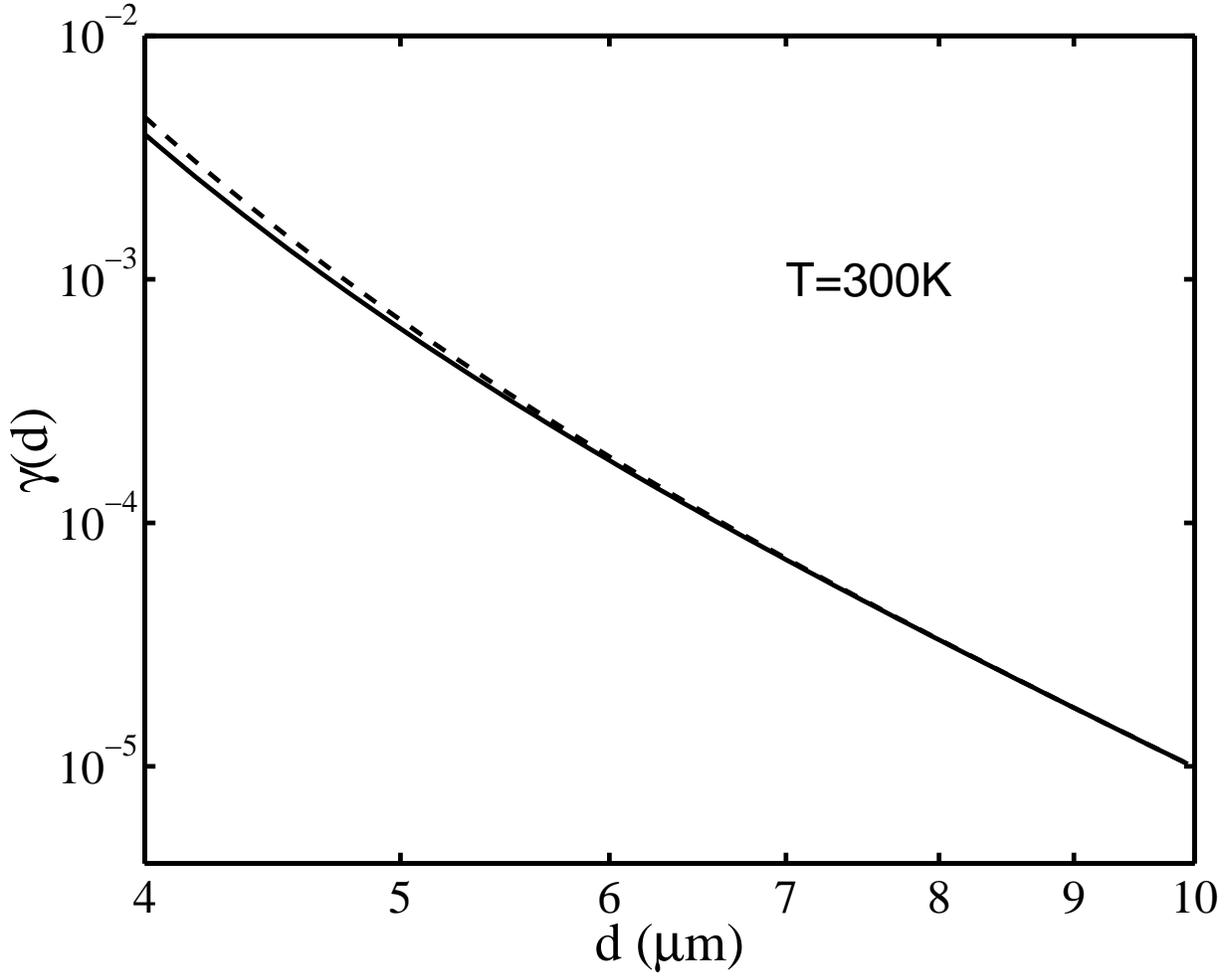}
\caption{\footnotesize Fractional  frequency shift (\ref{gamma}) of the center of mass oscillation (solid line) as a function of the distance from the surface, compared with the prediction of the static approximation (dashed line). 
The curves were obtained for a sapphire substrate at $300 K$ and for $^{87}$Rb atoms in the condensate. The unperturbed frequency was
$\omega_z/2\pi=220 Hz$ while the radius of the condensate was $R_z=2.5 \mu m$.}\label{fig:5}
\end{center}
\end{figure}
\begin{figure}[ptb]
\begin{center}
\includegraphics[width=1.0\textwidth]{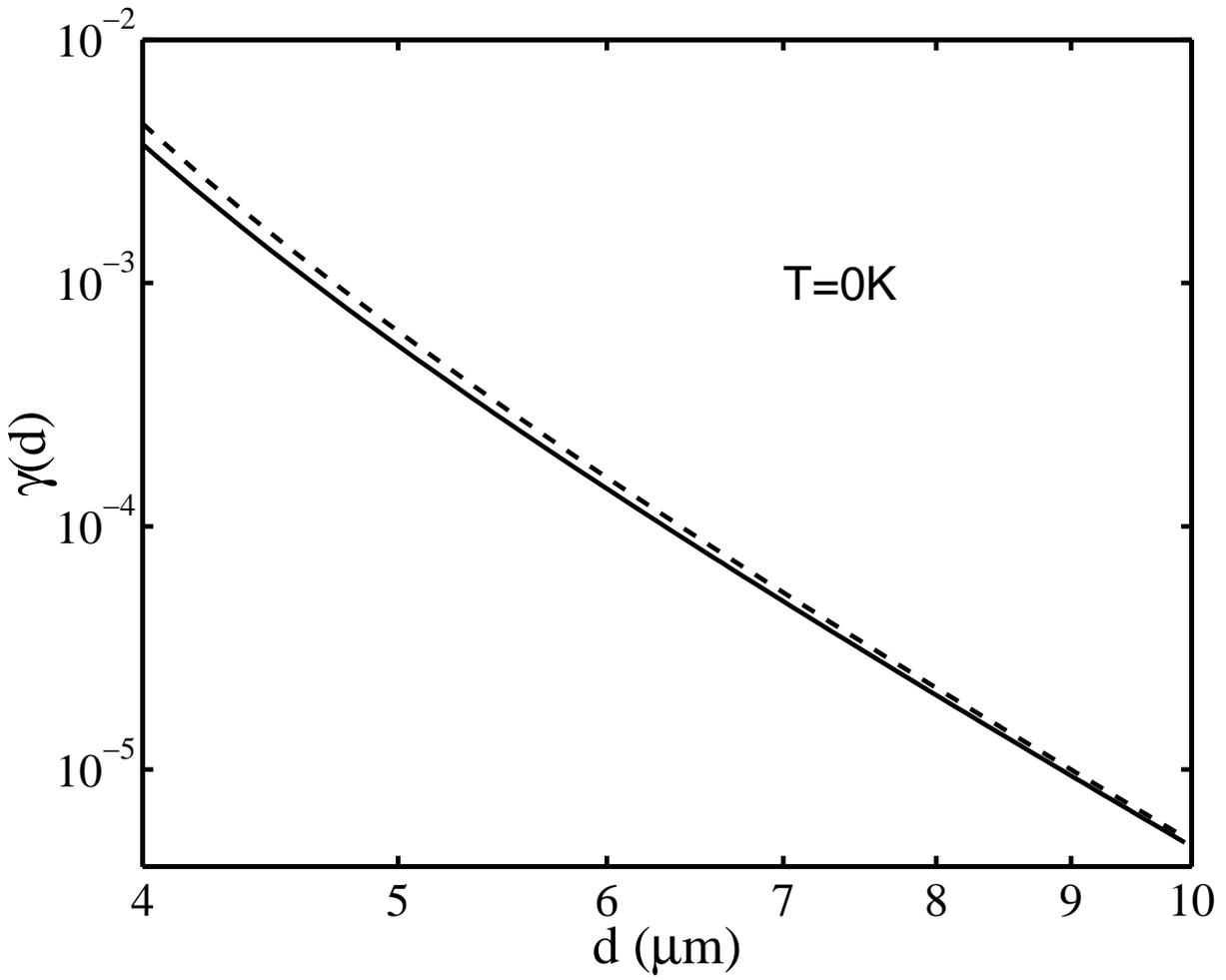}
\caption{\footnotesize Same as in fig.5, at $T=0 K$.}\label{fig:6}
\end{center}
\end{figure}
\begin{figure}[ptb]
\begin{center}
\includegraphics[width=1.0\textwidth]{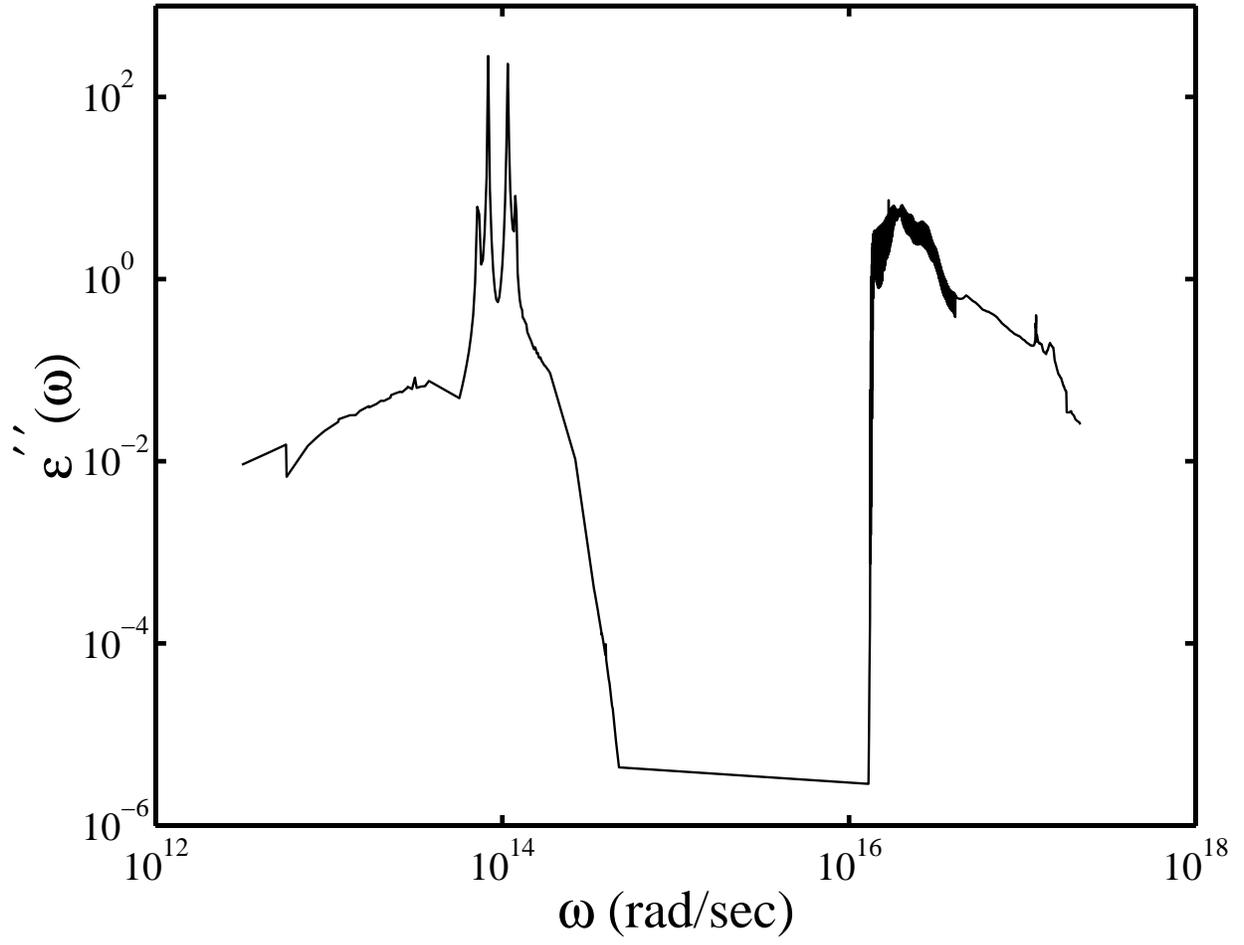}
\caption{\footnotesize Immaginary part of the sapphire dielectric function 
$\varepsilon^{\prime \prime}(\omega)$ 
 on the real axis of frequences.}\label{fig:7}
\end{center}
\end{figure}
\begin{figure}[ptb]
\begin{center}
\includegraphics[width=1.0\textwidth]{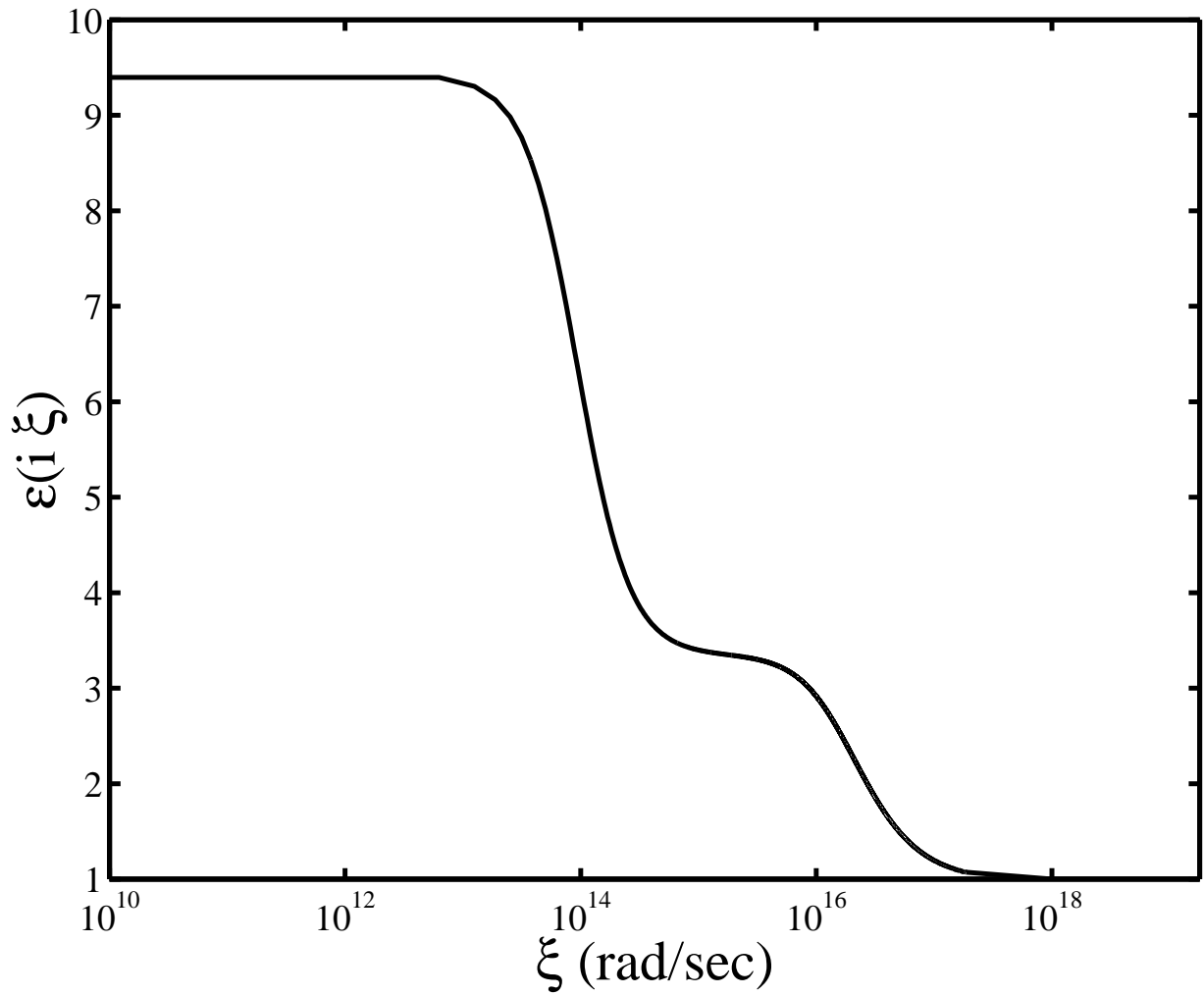}
\caption{\footnotesize Sapphire dielectric function $\varepsilon(i\;\xi)$ evaluated on the imaginary axis.}\label{fig:8}
\end{center}
\end{figure}
\begin{figure}[ptb]
\begin{center}
\includegraphics[width=1.0\textwidth]{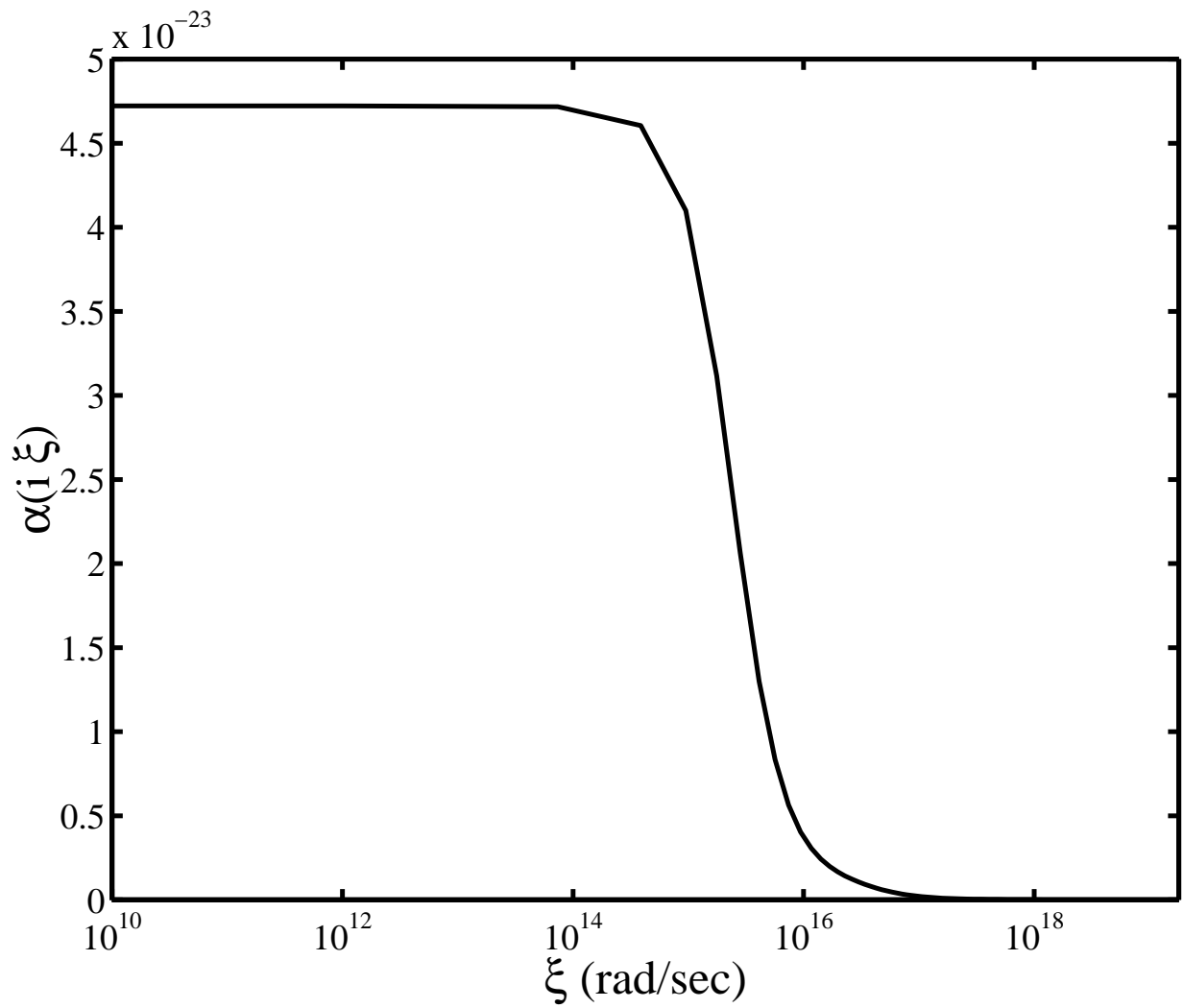}
\caption{\footnotesize Polarizability function $\alpha(i\;\xi)$ (in $cm^3$) for $^{87}$Rb atoms evaluated on the imaginary axis.}\label{fig:9}
\end{center}
\end{figure}
\begin{figure}[ptb]
\begin{center}
\includegraphics[width=1.0\textwidth]{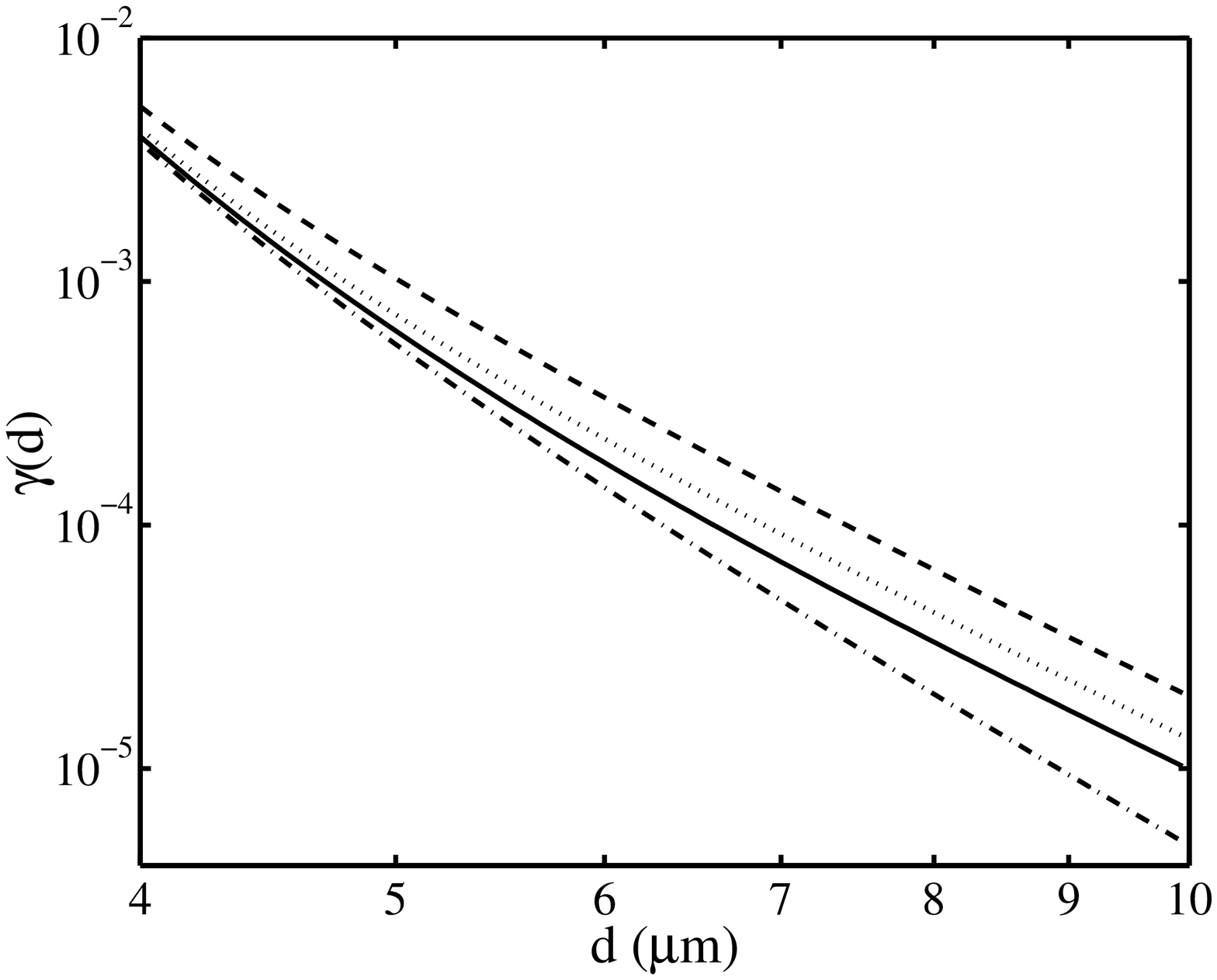}
\caption{\footnotesize Relative frequency shift (\ref{gamma}) of the center of mass oscillation calculated  at $T=0K$ (dash-dotted line), $T=300K$ (solid
line), $T=400K$ (dotted line) and at $T=600K$ (dashed line). The curves were obtained for a sapphire substrate  and for $^{87}$Rb atoms in the condensate. The unperturbed frequency was
$\omega_z/2\pi=220 Hz$ while the radius of the condensate was $R_z=2.5 \mu m$.}\label{fig:10}
\end{center}
\end{figure}


\begin{thebibliography}{99}

\bibitem{babb} J. F. Babb, G. L. Klimchitskaya, V. M. Mostepanenko , preprint quant-phys/0405163.

\bibitem{milton} K.A. Milton, preprint hep-th/0406024. 

\bibitem{capasso} H.B.~Chan et al., Phys. Rev. Lett.,{\bf 67}, 211801 (2001).

\bibitem{dimopulos}S.~Dimopoulos and A.A.~Geraci, Phys. Rev. D, {\bf 68}, 124021 (2003).

\bibitem{24} M.~Bordag, U.~Mohideen, and V.M.~Mostepanenko, Phys. Rep. {\bf 353}, 1 (2001).

\bibitem{4} Y.~Lin, I.~Teper, C.~Chin, and V.~Vuletic, Phys. Rev. Lett. {\bf 92}, 050404 (2004).

\bibitem{26} A.E.~Leanhardt et al., Phys. Rev. Lett., {\bf 90}, 100404 (2003).

\bibitem{27}  D.M.~Harber, J.M.~McGuirk, J.M.~Obrecht, and E.A.~Cornell, J. Low. Temp. Phys., {\bf 133}, 229 (2003);  J.M.~McGuirk, D.M.~Harber, J.M.~Obrecht, and E.A.~Cornell, Preprint cond-mat/0403254. 

\bibitem{london} F.~London, Z. Phys., {\bf 60}, 491 (1930).

\bibitem{CP} H.B.G.~Casimir and D.~Polder, Phys. Rev, {\bf 73}, 360 (1948).

\bibitem{additivity} It is worth pointing out that the interaction between macroscopic bodies is not in general the result of a simple summation of the interatomic force. Additivity   is  guaranteed only in the case of rarefied gases. 

\bibitem{lifshitz} E.M.~Lifshitz, Sov. Phys. JETP, {\bf 3}, 977 (1957).

\bibitem{LP} I.E.~Dzyaloshinskii, E.M.~Lifshitz and L.P.~Pitaevskii, Advances in Physics, Part 38, 165 (1961).

\bibitem{rmp}  F. Dalfovo, S. Giorgini, L.P. Pitaevskii, and S. Stringari,
Rev. Mod. Phys. {\bf 71}, 463 (1999).

\bibitem{leggett} A.J.~Leggett, Rev. Mod. Phys., {\bf 73}, 307 (2001).

\bibitem{CPHS}  C.J.~Pethick and H.~Smith, {\it Bose-Einstein condensation in
dilute Bose gases}. Cambridge University Press, 2002. 

\bibitem{LPSS} L.P.~Pitaevskii and S.~Stringari, {\it Bose-Einstein condensation}, Oxford University Press, 2003.

\bibitem{mitint}  M.R.~Andrews et al., Science, {\bf 275}, 637 (1997).

\bibitem{kasevich}  C.~Orzel et al., Science, {\bf 291}, 2386 (2001).

\bibitem{SS}  S.~Stringari, Phys. Rev. Lett., {\bf 77}, 2360 (1996).

\bibitem{jila96} D.S.~Jin et al., Phys. Rev. Lett., {\bf 77}, 420 (1996).

\bibitem{mit96}  D.M.~Stamper-Kurn et al., Phys. Rev. Lett., {\bf 81}, 500 (1998).

\bibitem{kohn}  W. Kohn, Phys. Rev. {\bf 123}, 1242 (1961).

\bibitem{sag} Due the external force generated by the wall, the center of mass position of the gas at 
rest does not 
coincide with the center of the harmonic trap. This sag effect is however very small  and can be safely  ignored. 

   
\bibitem{fermi} In the case of  an  ideal  degenerate Fermi gas confined in a harmonic trap one instead finds the result   $n_0^z(z)=(16/5\pi R_z)(1-z^2/R_z^2)^{5/2}$, where $R_z$ is the corresponding Thomas-Fermi radius, determined  by the Fermi energy $E_F=\hbar\omega_{ho}(6N)^{1/3}$ according to the formula   $E_F=(1/2)m\omega_z^2R_z^2$. Here $N$ is the number of trapped atoms of a single spin species.

\bibitem{Tneglected} Temperature can still affect the value of the dielectric and polarizability functions, but this effect is small  and has been ignored in the present work.

\bibitem{opposite} In this paper we will not discuss  the opposite limit $\lambda_T \ll \lambda_{opt}$ which is never reached in realistic conditions. 

\bibitem{abrikosova} B.V. Deryagin, I.I. Abrikosova and E.M. Lifshitz, Quart. Rev. Chem. Soc. {\bf 10}, 295 (1958); Uspekhi Fiz. Nauk {\bf 64}, 493 (1958).  

\bibitem{babbprivate} J.F. Babb, private communication.

\bibitem{expsapphire} W.J.~Tropf and M.E.~Thomas, In: {\it Handbook of optical constants of solids}, Vol. III, Academic Press, 1998.

\bibitem{lev} L.P.~Pitaevskii, Phys. Lett. A, {\bf 221}, 14 (1996); Yu. Kagan, E.L. Surkov and G.V. Shlyapnikov, Phys. Rev. A {\bf 54}, R1753 (1996).

\bibitem{dalibard} F.~Chevy et al, Phys. Rev. Lett., {\bf 88}, 250402 (2002).

 
\end{thebibliography}
\end{document}